\documentclass[12pt,twoside]{article}

\usepackage[mathscr]{eucal}
\usepackage{amsmath,amsfonts,amssymb,amsthm}
\usepackage{times}
\usepackage{srcltx}
\usepackage{pdfsync}
\usepackage{hyperref}

\advance\textheight\topskip
\numberwithin{equation}{section} \makeatletter
\@addtoreset{equation}{section}

\newtheorem{prop}{Proposition}[section]

\renewcommand{\tilde}{\widetilde}
\renewcommand{\hat}{\widehat}

\newcommand{\bref}[1]{\textbf{\ref{#1}}}

\renewcommand{\mod}{\,\rm mod \,}
\newcommand{\p}[1]{|#1|}
\newcommand{\gh}[1]{\mathrm{gh}(#1)}

\newcommand{\dd}{\partial}
\renewcommand{\d}{\partial}

\renewcommand{\geq}{\,{\geqslant}\,}
\renewcommand{\leq}{\,{\leqslant}\,}

\newcommand{\inner}[2]{\langle #1{,}\,#2\rangle}
\newcommand{\binner}[2]{%
  {\langle}\kern-4.15pt{\langle}#1{,}\,#2{\rangle}\kern-4.15pt{\rangle}}
\newcommand{\commut}[2]{[#1{,}\,#2]}

\newcommand{\pb}[2]{\left\{{}#1{},{}#2{}\right\}}
\newcommand{\ab}[2]{\big(#1,#2\big)}

\newcommand{\half}{\mathchoice{%
    \ffrac{1}{2}}{\frac{1}{2}}{\frac{1}{2}}{\frac{1}{2}}}

\newcommand{\ffrac}[2]{\raisebox{.5pt}%
  {\footnotesize$\displaystyle\frac{#1}{#2}$}\kern1pt}

\newcommand{\red}{\mathrm{red}}

\newcommand{\dl}[1]{\mathchoice{\ffrac{\dd}{\dd #1}}{\frac{\dd}{\dd
      #1}}{\ffrac{\dd}{\dd #1}}{\ffrac{\dd}{\dd #1}}}
\newcommand{\dr}[1]{\ffrac{{\overset{\leftarrow}{\partial}}}{ \partial #1}}
\newcommand{\drr}{\displaystyle{\overset{\leftarrow}{\partial}}}

\newcommand{\ddl}[2]{\ffrac{\dd #1}{\dd #2}}

\newcommand{\vdl}[1]{\ffrac{{\delta}}{\delta #1}}

\newcommand{\vddr}[2]{\ffrac{\delta^R #1}{\delta #2}}
\newcommand{\vddl}[2]{{\ffrac{\delta #1}{\delta #2}}}

\newcommand{\dlf}[1]{\mathchoice{\ffrac{\dd^F}{\dd #1}}{\frac{\dd^F}{\dd
      #1}}{\ffrac{\dd^F}{\dd #1}}{\ffrac{\dd^F}{\dd #1}}}
\newcommand{\drf}[1]{\ffrac{{\overset{\leftarrow}{\delta^F}}}{\dd #1}}

\newcommand{\manifold}[1]{\mathscr{#1}}

\newcommand{\manM}{\manifold{M}}

\def\cL{\mathcal{L}}

\def\cO{\mathcal{O}}
\def\cP{\mathcal{P}}

\newcommand{\F}{{F}}

\newcommand{\comment}[1]{}
\newcommand{\comp}[1]{{}}
\def\BGST{Barnich:2004cr}
\def\BGadS{Barnich:2006pc}

\def\Goff{Grigoriev:2006tt}

\def\AGT{Alkalaev:2008gi}
\def\GD{Grigoriev:1999qz}
\def\BGnlp{Barnich:2010sw}
\def\BG-Poincare{Barnich:2009jy}
\def\Fedosov-book{Fedosov:1996fu}

\begin{document}


\newpage

\begin{flushright}\small
FIAN-TD-2010-16
\end{flushright}

\begin{centering}

  \vspace{1cm}

  \textbf{\Large{Parent formulation at the Lagrangian level}}

  \vspace{1cm}

  {\large Maxim Grigoriev} 

\vspace{.7cm}

\begin{minipage}{.9\textwidth} \it \begin{center}
   Tamm Theory Department, Lebedev Physics
   Institute,\\ Leninsky prospect 53,  119991 Moscow, Russia\end{center}
\end{minipage}

\end{centering}

\vspace{1cm}

\begin{center}
  \begin{minipage}{.89\textwidth}
    \textsc{Abstract}. The recently proposed first-order parent
    formalism at the level of equations of motion is specialized to
    the case of Lagrangian systems. It is shown that for
    diffeomorphism-invariant theories the parent formulation takes the
    form of an AKSZ-type sigma model. The proposed formulation can be
    also seen as a Lagrangian version of the BV-BRST extension of the
    Vasiliev unfolded approach.  We also discuss its possible
    interpretation as a multidimensional generalization of the
    Hamiltonian BFV--BRST formalism.  The general construction is
    illustrated by examples of (parametrized) mechanics, relativistic particle,
    Yang--Mills theory, and gravity.

  \end{minipage}
\end{center}
\section{Introduction}

The Batalin--Vilkovisky (BV)
formalism~\cite{Batalin:1981jr,Batalin:1983wj} allows reformulating
nearly any gauge system as a universal BV theory that has an elegant
and unique form irrespective of the particular structure of the
starting point system. In so doing all the information about the
Lagrangian, gauge transformations, Noether identities and higher
structures of the gauge algebra are encoded in the BV master
action. This is achieved by introducing ghost fields and antifields in
such a way that the entire field-antifield space acquires an odd
Poisson bracket (the antibracket).  All the compatibility conditions like
gauge invariance of the action, reducibility relation and so on are
then encoded in the master equation which is merely equivalent to
requiring the BRST transformation to be nilpotent.

All the ingredients of the BV formalism can be naturally seen as geometric objects
defined on an abstract manifold and the BV formalism makes perfect sense in the
purely geometrical setting. In the context of local gauge field theory the
manifold in question has an extra structure: it is the space of suitable maps
(field histories) between the space-time and the target-space manifolds.
Moreover, all the ingredients such as the Lagrangian, gauge generators, structure
functions and so on are required to involve space-time derivatives of finite
order. In the BV formalism the locality is usually taken into
account~\cite{DuboisViolette:1985jb,Barnich:1995db,Piguet:1995er,Piguet:1995er} 
by approximating the space of field histories by the respective
jet bundle (see e.g.~\cite{Olver:1993,Anderson1991,Dickey:1991xa,vinogradov:2001} for a review on jet bundle approach). More technically, the formalism involves the total de Rham
differential along with the BRST differential so that the naive BRST complex
becomes a part of the appropriate bicomplex.

Although the jet space extension of the BV formalism has proved extremely useful
in studying, e.g., renormalization, anomalies, and consistent
deformations~\cite{DuboisViolette:1985jb,Piguet:1995er,Barnich:1993vg}
(see~\cite{Barnich:2000zw} for a review) it is not completely satisfactory
because the jet space approximation can be too restrictive. For instance, the
boundary dynamics is not captured in a straightforward way. In addition, the
jet space structures such as, e.g., generalized connections and curvatures
of~\cite{Brandt:1996mh,Brandt:1997iu,Brandt:2001tg} do not have a direct
dynamical meaning and are not manifestly realized in the formulation.

An interesting alternative to the jet space description of gauge
theories is the unfolded
formalism~\cite{Vasiliev:1988xc,Vasiliev:2005zu} developed in the
context of higher spin gauge theories. In this approach on-shell
independent derivatives of fields are treated as new independent
fields and the equations of motion are represented as a free
differential algebra (FDA)~\cite{Sullivan:1977fk}. The latter structure also underlies
somewhat related approaches to
supergravity~\cite{DAuria:1982nx,Fre:2008qw}.  It is within the
unfolded framework that the interacting theory of higher spin fields
on the AdS space has been derived~\cite{Vasiliev:1988sa,Vasiliev:1992av,Vasiliev:2003ev}.
The unfolded approach is also a powerful tool in studying gauge field theories invariant
under one or another space-time symmetry algebras~\cite{Vasiliev:2001zy,Shaynkman:2004vu}.

At the level of equations of motion the relation between the BV
formalism and the unfolded approach was established in~\cite{\BGST}
(see also~\cite{\BGadS,Barnich:2005ru}) for linear systems and in
\cite{\BGnlp} in the general case by constructing the so-called parent
formulation such that both the BV and the unfolded formulation can be
arrived at via straightforward reductions. The parent formulation
itself or some of its extensions can be considered as a new
formulation generalizing and unifying both the BV and the unfolded
formulation at the level of equations of motion.  Moreover, it is the
parent formulation that gives a systematic way to construct (and proves
the existence of) the unfolded form of a given theory.

In this paper we specialize the parent formulation to the case of
Lagrangian systems giving a parent extension of the BV formalism.
In particular, we identify the precise set of fields and antifields,
prescribe the antibracket and construct the master action satisfying
the classical master equation. We show that for diffeomorphism-invariant theories the parent formulation is a
sigma model of Alexandrov--Kontsevich--Schwartz--Zaboronsky (AKSZ) type~\cite{Alexandrov:1997kv} (see also
\cite{Cattaneo:1999fm,%
  Grigoriev:1999qz,Batalin:2001fc,Batalin:2001fh,
  Cattaneo:2001ys,Park:2000au,Roytenberg:2002nu,Kazinski:2005eb,Ikeda:2006wd,
  Bonechi:2009kx,Barnich:2009jy} for further developments
and applications of AKSZ-type sigma models) for which the target space is the BV jet space
of the starting point system while the starting point Lagrangian plays
the role of a potential.

\section{Parent Lagrangian}
\subsection{Preliminaries}
Suppose we are given a regular local Lagrangian gauge field
theory. Within the BV formalism the theory is defined by the master
action $S[\psi,\psi^*]$, where $\psi^A,\psi_A^*$ are fields and
antifields. The space of fields and antifields carry an integer ghost
degree $\gh{\cdot}$ such that fields of the theory are those $\psi^A$
with $\gh{\psi^A}=0$ while the remaining $\psi^A$-s are ghost fields,
ghosts for ghosts, and so on, and carry positive ghost degrees.  The
master action $S$ carries vanishing ghost degree and satisfies the
master equation
\begin{equation}
 \ab{S}{S}=0\,,
\end{equation} 
with respect to the antibracket defined by
\begin{equation}
 \ab{\psi^A(x)}{\psi^*_A(x^\prime)}=\delta^A_B\delta^{(n)}(x-x^\prime)\,, \quad \ab{\psi^A(x)}{\psi^B(x^\prime)}=\ab{\psi^*_A(x)}{\psi^*_B(x^\prime)}=0\,,
\end{equation}
where $x^\mu, \mu=0,\ldots, n-1$ denote space-time coordinates.
The ghost numbers and the Grassmann parities of the antifields are
determined by those of the fields through
$\gh{\psi^*_A}=-1-\gh{\psi^A}$ and $\p{\psi^*_A}=\p{\psi^A}+1\mod 2$
so that the antibracket is Grassmann odd and carries ghost degree $1$.

We restrict ourselves to the case of theories with closed algebra. For such theories $S[\psi,\psi^*]$
can be chosen at most linear in antifields. More precisely, $S$ can be taken as
\begin{equation}
\label{S-orig}
 S=\int d^n x L_0[\psi]+\int d^n x\, \psi^*_A (\gamma \psi^A)  \,,
\end{equation} 
where $\gamma$ is a gauge part of the complete BRST differential $s$
and $L_0[\psi]$ is the Lagrangian.  In our case, $\gamma$ is nilpotent
and enters the complete BRST differential $s=\ab{\cdot}{S}$ as
$s=\delta+\gamma$. Here, $\delta$ is the Koszul--Tate term
implementing the equations of motion determined by $L_0$ and their
reducibility relations. Note that in general $\gamma$ is nilpotent
only modulo equations of motion and $s=\delta+\gamma+\ldots\,$, where
dots refer to terms originating from the terms in $S$ of the second
and higher orders in $\psi^*_A$.

We first recall the construction~\cite{\BGnlp} of the parent
theory at the level of the equations of motion.  In the present
context it is convenient to concentrate on the gauge structure encoded
in $\gamma$ and temporarily disregard the actual equations of motion
implemented through $\delta$ and the antifields $\psi^*_A$.  This
corresponds to the off--shell version of the parent formulation
in~\cite{\BGnlp}.  The extended set of fields (including ghost fields
etc.) is given by $\psi^A_{(\lambda)[\nu]}$, where $(\lambda)$ denotes
a symmetric multi-index and $[\nu]$ an antisymmetric one. Introducing
bosonic variables $y^\lambda$ and fermionic variables $\theta^\nu$, all
the fields can be packed into the generating function
\begin{equation}
\begin{split}
\label{decomposition}
\tilde\psi^A(x,y,\theta)=\sum_{k,l\geq 0} \frac{1}{k!l!}
\theta^{\nu_l}...\,\theta^{\nu_1}y^{\lambda_k}...\, y^{\lambda_1}\psi^A_{\lambda_1...\,
  \lambda_k|\nu_1...\,\nu_l}(x)
\equiv \theta^{(\nu)} y^{(\lambda)}\psi^A_{(\lambda)[\nu]}(x)\,.
\end{split}
\end{equation}
The ghost degrees of the component fields are determined by the ghost
degree of $\psi^A$ if one prescribes $\gh{y^\lambda}=0$ and
$\gh{\theta^\nu}=1$. For instance, $\gh{\psi^A_{(\lambda)|\nu}}=\gh{\psi^A}-1$.
In what follows we also use the condensed
notation $\psi^\alpha$ for all the fields so that $\alpha$ stands for
$A,(\mu),[\nu]$ and ranges over an infinite but countable set.  The
lowest component $\psi^A_{()[\,]}$ is identified with $\psi^A$. Fields $\psi^A_{(\lambda)[\nu]}$
are refereed to as $\theta$ and $y$-derivatives (or descendants) of $\psi^A$.

We need to introduce some useful operations on the space of fields of the parent theory.  Given a differential
operator $\cO$ on the space of $y,\theta$ and $x$ we associate a functional vector field $\cO^F$ on the space of
fields $\psi^A_{(\lambda)[\nu]} (x)$ according to  (see ~\cite{\BGnlp} for more details)
\begin{equation}
\label{op2F}
 \cO^F(\tilde\psi^A)=(-1)^{\p{A}\p{\cO}}\cO\tilde\psi^A \,,
\end{equation} 
where $\cO^F$ is assumed to act from the right. Here, $\cO$ 
acts on $y,\theta,x$ while $ \cO^\F$ acts on the space of fields 
$\psi^A_{(\lambda)[\nu]}(x)$. 
 Relation~\eqref{op2F} is compatible with the commutator in the sense that $(\commut{\cO_1}{\cO_2})^F=\commut{\cO_1^F}{\cO_2^F}$. To fit with the usual conventions for the master action (see, e.g.,~\cite{HT-book}) we have exchanged the left and right action with respect to~\cite{\BGnlp}.
Using~\eqref{op2F} one defines $d^F,\sigma^F,\dlf{y^\mu},\dlf{\theta^\mu}$ associated to
$\sigma=\theta^\mu\dl{y^\mu}$, $d=\theta^\mu\dl{x^\mu}$, $\dl{y^\mu}$, and $\dl{\theta^\mu}$. In what follows we need some explicit relations:
\begin{equation}
\begin{gathered}
 \dlf{\theta^\nu}\psi^A=(-1)^{\p{A}}\psi^A_{()\nu}\,,  \quad 
\dlf{y^\nu}\psi^A=\psi^A_{\nu[]}\,, \quad d^F\psi^A_{(\lambda)[\,]}=\sigma^F\psi^A_{(\lambda)[\,]}=0\,,\\ 
d^F\psi^A_{()\nu}=(-1)^{\p{A}} \d_\nu \psi^A\,,
\quad   \sigma^F\psi^A_{()\nu}=(-1)^{\p{A}} \psi^A_{\nu[\,]}\,.
\end{gathered}
\end{equation} 

We often employ the language of jet spaces (see, e.g.,~\cite{Anderson1991,Olver:1993}) and hence replace
the space of field histories $\psi^\alpha(x)$ by the respective jet space with coordinates $x^\mu$, $\psi^\alpha$,
and all $x$-derivatives $\psi^\alpha_{(\mu)}$. We also use $\d_\mu$ to denote the total derivative:
\begin{equation}
 \d_\mu=\dl{x^\mu}+\psi^\alpha_{\mu}\dl{\psi^\alpha}+\psi^\alpha_{\mu\mu_1}\dl{\psi^\alpha_{\mu_1}}+\ldots\,.
\end{equation}
Functional vector fields defined by~\eqref{op2F} can be also seen as vector fields on the jet space.

The gauge part $\gamma$ of the BRST differential can then be naturally
seen as acting on the space with coordinates $x^\mu,\psi^\alpha_{(\mu)}$.  This is
achieved as follows: for $\psi^A$ one defines $\bar\gamma
\psi^A=\gamma \psi^A$, where the derivatives $\d_{(\mu)}\psi^A$ in the
HRS are replaced by $\psi^A_{(\mu)[\,]}$. The action of $\bar\gamma$
on coordinates $\psi^A_{(\lambda)[\,]}$ is uniquely determined by
requiring $\commut{\dl{x^\mu}+\dlf{y^\mu}}{\bar\gamma}=0$. Finally the
action on $\theta$-derivatives $\psi^A_{(\lambda)[\nu]}$ and
$x$-derivatives of all the fields is obtained by the usual
prolongation $\commut{\d_\mu}{\bar\gamma}=\commut{\dlf{\theta^\nu}}{\bar\gamma}=0$.

Finally, the BRST differential of the parent theory is given by~\cite{\BGnlp}
\begin{equation}
\gamma^P=d^F-\sigma^F+\bar\gamma\,. 
\end{equation} 
It was shown in~\cite{\BGnlp} that the parent formulation is
equivalent to the starting point one via elimination of generalized
auxiliary fields (see Section~\bref{sec:equiv} for the definition
and~\cite{Dresse:1990dj,\BGST} for details on this notion of
equivalence).

\subsection{Parent master action}
\label{sec:parent-action}

To simplify the exposition, we assume for the moment that the starting point Lagrangian $L_0[\psi]$ is strictly gauge invariant so that $\gamma L_0=0$. The general case where $L_0$ is gauge invariant modulo a total derivative is considered next.

Associated to each field $\psi^\alpha$ we introduce an antifield $\Lambda_\alpha$ or in components $\Lambda_A^{(\mu)[\nu]}$ and postulate the usual antibracket, ghost number and Grassmann parity assignments:
\begin{equation}
\label{ab}
\begin{gathered}
 \ab{\psi^\alpha(x)}{\Lambda_\beta(x^\prime)}_P=\delta^\alpha_\beta\delta^{(n)}(x-x^\prime),\\
 \gh{\Lambda_\alpha}=-\gh{\psi_\alpha}-1,\quad \p{\Lambda_\alpha}=\p{\psi_\alpha}+1\mod 2  \,.
\end{gathered}
\end{equation} 
Consider then the following functional
\begin{equation}
\label{main}
S^P=\int d^n x \,\left(\Lambda_\alpha(d^\F-\sigma^\F+\bar\gamma)\psi^\alpha + L_0(\psi^A_{(\lambda)[\,]},x)\right)\,,
\end{equation} 
where $L_0(\psi^A_{(\lambda)[\,]},x)$ is the starting point Lagrangian
in which derivatives $\d_{(\mu)}\psi^A$ are replaced with
$\psi^A_{(\mu)[\,]}$.  Because space-time derivatives enter only
through $d^\F$ this action is a first-order one.
\begin{prop}
$S^P$ satisfies the master equation along with the usual ghost number and Grassmann parity assignments
\begin{equation}
 \ab{S^P}{S^P}_P=0\,, \qquad \gh{S^P}=0\,,\quad \p{S^P}=0\,,
\end{equation}
and hence can be considered a BV master action of a gauge
field theory.
\end{prop}
\begin{proof}
  It is useful to work in terms of integrands (understood modulo total
  derivatives). Let
  $K=\Lambda_\alpha(d^\F-\sigma^F+\bar\gamma)\psi^\alpha$ and $L_0$ be
  the integrands of respectively the first and the second terms
  in~\eqref{main}.  The equation $\ab{K}{K}_P=0$ is just a consequence
  of the nilpotency of the vector field
  $d^\F-\sigma^F+\bar\gamma$. $\ab{L_0}{L_0}_P=0$ is obvious because
  $L_0$ is independent of the antifields. Finally, nonvanishing
  contributions to $\ab{L_0}{K}_P$ can only originate from terms in
  $K$ involving $\Lambda_A^{(\mu)[0]}$.  But
  $(d^\F-\sigma^F)\psi^A_{(\mu)[0]}=0$ so that
  $\ab{L_0}{K}_P=\ab{L_0}{\Lambda_A^{(\mu)[0]}\bar\gamma\psi^A_{(\mu)[0]}}_P=0$
  as a consequence of $\gamma L_0=0$.
\end{proof}

The number of fields entering master action~\eqref{main} is
infinite. This complicates the analysis and makes ambiguous the interpretation
of~\eqref{main} as a BV action of a local gauge field theory. Fortunately, it turns out that the action
can be consistently truncated to the one involving only finitely many
fields and finitely many terms. To see this, we consider the degree
$N_{\d_y}+N_{\d_\theta}$, called truncation degree, where
\begin{equation}
 N_{\d_y}=\sum_{l\geq 0}l\psi^A_{\lambda_1\ldots\lambda_l[\nu]}\dl{\psi^A_{\lambda_1\ldots\lambda_l[\nu]}}\,,
\qquad
N_{\d_\theta}=\sum_{l\geq 0}l\psi^A_{(\lambda)\nu_1\ldots \nu_l}\dl{\psi^A_{(\lambda)\nu_1\ldots \nu_l}}\,.
\end{equation} 
To construct the truncated theory let us fix integer $M$ which is sufficiently high with respect to
the degree in $x$-derivatives of the starting point Lagrangian and BRST differential $\gamma$. For a given $m>M$
coordinates $\psi^A_{\lambda_1\ldots\lambda_k|\nu_l\ldots \nu_l}$ with $k+l=m$ can be replaced with coordinates
$w^{a_m},v^{a_m}$ such that $\sigma^Fw^{a_m}=v^{a_m}$ because all the coordinates except $\psi_{()[]}$
are contractible pairs for $\sigma^F$ as a consequence of Poincar\'e Lemma. Moreover, it was shown in~\cite{\BGnlp}
that equations $(d^F-\sigma^F+\bar\gamma)w^{a_m}$ can be algebraically solved for $v^{a_m}$ at $w^{a_m}=0$. 

The truncated formulation is obtained by imposing the following constraints:
\begin{equation}
 w^{a_m}=0\,,\quad (d^F-\sigma^F+\bar\gamma)w^{a_m}=0\,, \quad w^*_{a_m}=0\,, \quad v^*_{a_m}=0\,,\qquad m>M\,.
\end{equation} 
These constraints are equivalent to algebraic and moreover are second class constraints in the antibracket sense.
This guarantees that truncated master action $S^P_{(M)}$ satisfies the master equation. Moreover, $S^P_{(M)}$ has the following structure
\begin{equation}
\label{main-tr}
S^P_{(M)}=\int d^n x \,
\left(\sum_{k+l\leq M}\Lambda_{\alpha_{k,l}}(d^\F-\sigma^\F+\tilde{\bar\gamma})\psi^{\alpha_{k,l}}
+ L_0(\psi^A_{(\lambda)[\,]},x)\right)\,,
\end{equation} 
where $\psi^{\alpha_{k,l}},\Lambda_{\alpha_{k,l}}$ denote $\psi^A_{\lambda_1\ldots\lambda_k|\nu_l\ldots \nu_l}$ 
and their conjugate antifields. Note that thanks to the above constraints the differential $\bar\gamma$ is
replaced by its modification $\tilde{\bar\gamma}$. The modification actually affects the action on higher
degree fields only: $\tilde{\bar\gamma}\psi^{\alpha_{k,l}}={\bar\gamma}\psi^{\alpha_{k,l}}$ for $k+l<M-T$,
where $T$ denotes the total degree of initial $\gamma$ in space-time derivatives.  This implies that the parent action
and its truncation coincide up to terms involving variables of degree higher than $M-T$. Because $T$ is fixed
and $M$ is arbitrary but finite one can consider $S^P$ as a sort of limit of $S^P_{(M)}$ as
$M\to \infty$.\footnote{
Note that the above truncation is far from being unique. In concrete examples one or another equivalent choice
can be useful. For instance for linear theories one can simply put to zero all the fields with degree
$N_{\d_y}+N_{\d_\theta}-T{\rm gh_T}$ higher than a truncation bound. Here ${\rm gh_T}$ is the target space ghost degree defined through ${\rm gh_T}(\psi^A_{(\lambda)[\nu]})=\gh{\psi^A}$.}

This observation gives the parent theory the following interpretation: this is the theory determined by 
$S^P_{(M)}$ where the truncation bound $M$ is chosen high enough but finite. In fact, it is even useful
not to fix the truncation bound and work as if all necessary fields were present.
This interpretation makes sense because the equations of motion, gauge symmetries etc. for fields of truncation degree
less than $M-T$ do not depend on $M$. Here and in what follows we assume that $\gamma$ and $L_0$ involve derivatives
up to a finite order and the ghost degree of fields $\psi^A$ is also finite. In particular, this is necessary for the above truncation to exist.

In what follows we refer to the local gauge field theory determined by
$S^P$ (or its generalizations considered below) as the parent
formulation. According to the principles of the BV formalism the
fields of the parent formulation are those fields among
$\psi^\alpha,\Lambda_\alpha$ that have the vanishing ghost degree. The
respective classical action $S^P_0$ is obtained from $S^P$ by putting all the
fields of a nonvanishing ghost degree to zero.  Gauge transformations
for the fields are then read off from the complete BRST differential
$s^P=\ab{\cdot}{S^P}$ by $ \delta \phi^i=s^P\phi^i $, where in the
Right-Hand Side we put all the fields of ghost degrees different from
$0,1$ to zero and replace degree-$1$ fields with gauge parameters.

\comment{In the paragraph above it probably makes sense to decompose $S^p$ in terms of antifields and ghost fields
explicitly.}

It turns out that the parent formulation determined by $S^P$ is equivalent to
the starting point theory determined by $S$ through the elimination of
generalized auxiliary fields. It is then a BV master action for the parent
theory of~\cite{\BGnlp} in the case where the starting point theory is
Lagrangian (recall also that $\gamma L_0=0$ and the gauge algebra is closed in
our setting). Moreover, $S^P$ is a proper solution to the master equation
provided the starting point $S$ is a proper one. In the rest of the paper we
extend the construction to generic gauge theories, identify the structure of the
parent formulation for diffeomorphism-invariant theories, prove the equivalence
to the starting point theory, and illustrate the constructions by concrete
examples. 

\subsection{Equivalence proof}
\label{sec:equiv}

According to the definition from~\cite{Dresse:1990dj}
fields $\chi^i,\chi^*_i$ are generalized auxiliary fields
for the master action $S$ if they are canonically conjugate in the antibracket
and equations $\vddl{S}{\chi^i}\big|_{\chi^*_i=0}=0$ can be algebraically solved for $\chi^i$.
\begin{prop}
  The BV formulation determined by $S^P,\ab{\cdot}{\cdot}_P$ and the
  starting point theory $S,\ab{\cdot}{\cdot}$ are equivalent via
  elimination of generalized auxiliary fields.
\end{prop}
\begin{proof}
  All the fields $\psi^A_{(\lambda)[\nu]}$ save for
  $\psi^A=\psi^A_{(\,)[\,]}$ can be grouped into two sets $w^a$ and $v^b$
  in such a way that $\sigma^\F w^a=v^a$. The set of fields and antifields can
  then be split as $\psi^A,\Lambda_A,w^a,v^a,w^*_a,v^*_a$.  Let us
  show that $v^a,w^a,v_a^*,w_a^*$ are generalized auxiliary fields. More precisely, as
  $\chi^i$ and $\chi^*_i$ we take respectively $v^a,w^*_a$ and $v^*_a,w^a$.

Varying first with respect to $w^*_a$ and putting $v^*,w$ to zero,
we find
\begin{equation}
 \left[(d^F-\sigma^F+\bar\gamma)w^a\right]|_{w=0}=0 \quad \Leftrightarrow \quad
 v^a=\left[(d^F+\bar\gamma)w^a\right]|_{w=0}\,.
\end{equation} 
It is almost clear from the last formula that it can be solved for
$v^a$.  The detailed proof uses the extra degrees (ghost degree and
$N_{\d_\theta}$) and was given in detail in~\cite{\BGnlp}.  In
particular, one finds that all $v^a$ vanish except for
$\psi^A_{(\lambda)[\,]}$.  If the theory is not truncated then
$\psi^A_{(\lambda)[\,]}=\d_{(\lambda)}\psi^A$. For the truncated theory
this is only true for lower order derivatives~\cite{\BGnlp}.  However,
if the truncation degree is high enough this does not affect the
reduced action because $L_{0}$ involves $y$-derivatives of bounded
order.

Varying then with respect to $v^a$ and putting $v^*,w$ to zero gives:
\begin{equation}
 w^*_a=\vddr{}{v^a}\left[w^*_b(d^F+\bar\gamma)w^b +\Lambda_A\bar\gamma\psi^A+L_0\right]\Big|_{w=0}\,.
\end{equation} 
The second and the third terms cannot spoil the solvability with
respect to $w^*_a$ because they do not involve $w^*_a$. To see that
this is also true for the first term, we use the following
modification of the truncation degree:
$N_{\d_y}+N_{\d_\theta}- (T+1){\rm gh_T}$.  In the linear order, we
then find that $((d^F+\bar\gamma)w^b)|_{w=0}$ can only involve
variables $v$ of the degree lower than that of $w^b$. It follows that
$(\vdl{v^a}(w^*_b(d^F+\bar\gamma)w^b)|_{w=0}$ can only involve
$w^*$-variables of degree higher then that of $w^*_a$. Because $S^P$
is assumed truncated and hence does not involve fields of sufficiently
high degree the equation can be solved order by order using the above degree
and the homogeneity in the fields.

Finally, putting to zero all $v^*_a,w^a$ as well as all $v^a$ except
$\psi^A_{(\mu)[\,]}=\d_{(\mu)}\psi^A$ the master action $S^P$ reduces
to
\begin{equation}
 \tilde S=S_0[\psi^A]+\Lambda_A \gamma \psi^A\,,
\end{equation} 
which is exactly the starting point master action~\eqref{S-orig} if one identifies $\Lambda_A$ with $\psi_A^*$.
\end{proof}

Now we are ready to discuss in some more details the truncation introduced in
the previous section. In particular, to relate it to elimination of generalized
auxiliary fields. To this end it is instructive to rewrite parent
action~\eqref{main} in the adapted coordinates
$\psi^A_{\lambda_1\ldots\lambda_k|\nu_l\ldots \nu_l}$ with $k+l \leq M$ and
$w^{a_m},{\bar v}^{a_m}=(d^F-\sigma^F+{\bar\gamma})w^{a_m}$ with $m>M$ and their
conjugate antifields.  Note that $\bar v^{a_m}$ replace coordinates
$v^{a_m}=\sigma^F w^{a_m}$. In terms of these coordinates the integrand of the
parent action takes the form 
\begin{equation}
\label{main-adapted}
\sum_{k+l\leq M}\Lambda_A^{\lambda_1\ldots\lambda_k|\nu_l\ldots \nu_l}
(d^\F-\sigma^\F+{\bar\gamma})\psi^A_{\lambda_1\ldots\lambda_k|\nu_l\ldots \nu_l}+\sum_{m>M}^\infty w^*_{a_m}{\bar v}^{a_m}+L_0(\psi^A_{(\lambda)[\,]},x)\,.
\end{equation} 
It is almost obvious from this representation that variables $w^{a_m},{\bar v}^{a_m},w^*_{a_m}, {\bar v}^*_{a_m}$
with $m>M^\prime$ for some $M^\prime\geq M$ are generalized auxiliary fields. Their elimination is noting but the truncation
at level $M^\prime$. In this representation it can look like the artificial truncation of the previous section is not needed as it can always be achieved by eliminating the above generalized auxiliary fields from the parent action. This is not the case, however, because the above change of variables contains $x$-derivatives (through $d^F$) and affects infinite number of 
coordinates so that it is not a strictly local operation. Indeed, taking for simplicity $\gamma=0$ one finds
that algebraic constraints $w^{a_m}={\bar v}^{a_m}=w^*_{a_m}={\bar v}^*_{a_m}=0$ in terms of original variables
involves any number of space-time derivatives. For instance, by these constraints $\psi^A_{\lambda_1\ldots \lambda_{M+k}[\,]}$ is expressed through $\d_{\lambda_1}\ldots \d_{\lambda_{k}}\psi^A_{\lambda_{k+1}\ldots\lambda_{k+M}[\,]}$.

\subsection{Generalization}

In order to allow for Lagrangians that are $\gamma$-closed only modulo
a total derivative we need some more technique. In the setting of the
starting point theory, we introduce the algebra of local forms
$\hat\Omega$ that are forms on $x$-space with values in local
functions. As a usual technical assumption we in addition exclude
field-independent forms from $\hat\Omega$. Local forms can be seen as
functions in the fields, their derivatives, the coordinates $x^\mu$,
and the fermionic variables $\theta^\mu$ standing for basic
differentials $dx^\mu$.  As is implied by the notation, the variables
$\theta^\mu$ are to be identified with the $\theta^\mu$ of the
previous sections.

In the usual local BRST cohomology considerations~(see,
e.g.,~\cite{Barnich:2000zw}) it is quite useful to employ the extended
BRST differential (recall that $\gamma$ acts from the right)
\begin{equation}
 \tilde\gamma=-d_H+\gamma\,, \qquad d_H = \overset{\leftarrow}\d_\mu \theta^\mu
\end{equation}
where $d_H$ is often refereed to as total de Rham differential.
For instance the ghost degree-$g$ cohomology of $\gamma$ in the space
of local functionals is in fact isomorphic to the total degree $g+n$ cohomology of
$\tilde \gamma$ in the space of local forms without
field-independent terms. The total degree extends ghost degree such that $\theta$
carries unit degree.

A particularly important representative of the local BRST cohomology
is the Lagrangian density itself~\footnote{Note that if instead of $\gamma$-cohomology one considers the cohomology of the complete
BRST operator $s=\gamma+\delta+\ldots$, a nontrivial Lagrangian can be a trivial representative of $s$-cohomology. For instance
this happens for free theories or pure gravity because the respective Lagrangians vanish on-shell.}.  It can be represented by a local
form $\hat L[\psi,x,\theta]$ of the total degree $n$ such that $\tilde
\gamma \hat L=0$.  The usual Lagrangian $L_0[\psi,x]$ enters $\hat
L$ as a coefficient of the volume form $\theta^0\ldots \theta^{n-1}$.  More
precisely $L_0[\psi,x]=\int d\theta^{n-1}\ldots d \theta^0 \hat
L[\psi,x,\theta]$ and $\tilde \gamma \hat L=0$ implies $\gamma
L_0=\d_{\mu}j_1^\mu$, $\gamma j_1^\mu=\d_\nu j_2^{\nu\mu}$, etc. with
some $j_k^{\mu_1\ldots \mu_k}$, $\gh{j_k}=k$. Note that because of the
above isomorphism any $L_0$ that is $\gamma$-closed modulo a total
derivative can be represented by such a $\tilde\gamma$-cocycle $\hat
L$. Obtaining $\hat L$ can be also seen as solving the respective
descent equation (see, e.g.,~\cite{Barnich:2000zw}) with
$\theta^1\ldots \theta^n L_0$ being the local form of maximal degree.

Representing the Lagrangian density through $\hat L$ we easily
generalize parent master action~\eqref{main} as
\begin{equation}
\label{action-mod}
 S^P=\int d^n x \,\left[ \Lambda_\alpha(d^\F-\sigma^\F+\bar\gamma)\psi^\alpha + \int
 d^n \theta \hat L(\tilde\psi^A_{(\lambda)},x,\theta)\right]\,,
\end{equation}  
where by a slight abuse of notation we have denoted $\tilde \psi^A_{(\lambda)}=
\sum \frac{1}{k!}\theta^{\nu_k}\ldots \theta^{\nu_1} \psi^A_{(\lambda)\nu_1\ldots \nu_k}\equiv\theta^{[\nu]}
\psi^A_{(\lambda)[\nu]}$.

Let us show that $S^P$ indeed satisfies the master equation modulo total
derivatives. The only nontrivial point is to check that $\gamma^P\int d^n\theta \hat L(\tilde\psi^A_{(\lambda)},x,\theta)$
is a total derivative. We first observe that
\begin{equation}
  \int d\theta^{n-1}\ldots d \theta^0 \hat L(\tilde\psi^A_{(\lambda)},x,\theta)=
\left[ \d^\theta_0\ldots \d^\theta_{n-1} \hat L(\psi^A_{(\lambda)},x,\theta)\right]\Big|_{\theta=0}\,,
\end{equation} 
where $\d^\theta_\mu=\dr{\theta^\mu}-\dlf{\theta^\mu}$ is a total right derivative with respect to $\theta^\mu$.
It is then useful to employ the extended parent differential~\cite{\BGnlp}:
\begin{equation}
{{(\tilde \gamma)}^P}=-(\dr{x^\mu} +
\dlf{y^\mu})\theta^\mu+d^F-\sigma^F+\bar \gamma\,,
\end{equation} 
which is nilpotent and satisfies 
${(\tilde \gamma)}^P|_{\theta=0}=\gamma^P$ and 
$\commut{\d^\theta_\mu}{(\tilde \gamma)^P}=-\d_\mu$. 

Using then 
$\commut{\d^\theta_\mu}{(\tilde \gamma)^P}=-\commut{\d^\theta_\mu}{d_H}$ gives
\begin{multline}
 \gamma^P\, \left[\d^\theta_0\ldots \d^\theta_{n-1} \hat L(\psi^A_{(\lambda)},x,\theta)\right] \Big|_{\theta=0}=
(-1)^{n} \left[ \d^\theta_0\ldots \d^\theta_{n-1} \, d_H \hat L(\psi^A_{(\lambda)},x,\theta)\right]\Big|_{\theta=0}=\\[6pt]
(-1)^{n} \int d^n\theta \,\, d_H \hat L(\tilde\psi^A_{(\lambda)},x,\theta)\,,
\end{multline}
so that the master equation is indeed satisfied modulo a total derivative.
Finally one can check that the equivalence proof of Section~\bref{sec:equiv} is
not affected by the extra terms in the parent Lagrangian. \comment{To be
checked!}

The structure of the parent formulation can be simplified by packing
the fields $\Lambda_A^{(\mu)[\nu]}$ into superfields
$\tilde\Lambda_A^{(\mu)}(\theta)$ such that $\Lambda_\alpha
\psi^\alpha=\Lambda_A^{(\mu)[\nu]}\psi^A_{(\mu)[\nu]}=(-1)^{n}\int d^n\theta
\tilde \Lambda_A^{(\mu)}\tilde \psi^A_{(\mu)}$. It is then useful to
employ the language of supergeometry.  Namely, consider a
supermanifold $\manM$ with coordinates being $\psi^A_{(\lambda)}$ and
$\Lambda^{(\lambda)}_A$, $\gh{\Lambda^{(\lambda)}_A}=-\gh{\psi^A_{(\lambda)}}+n-1$
and equipped with the (odd) Poisson bracket defined by
\begin{equation}
\label{brack-M}
\pb{\psi^A_{(\mu)}}{\Lambda^{(\nu)}_B}_\manM=\delta^A_B\delta_{(\mu)}^{(\nu)}\,.
\end{equation} 
The bracket carries ghost degree $1-n$ and the Grassmann parity $(1-n) \mod 2$.

We consider the function
\begin{equation}
  \label{Sm}
  S_\manM(\psi,\Lambda,x,\theta)=\Lambda^{(\mu)}_A \bar\gamma \psi^A_{(\mu)}+\hat L(\psi^A_{(\mu)},x,\theta)\,,
\end{equation} 
where as before $\hat L(\psi^A_{(\mu)},x,\theta)$ is obtained from $\hat L[\psi]$ by replacing $\d_{(\mu)}\psi^A$ with $\psi^A_{(\mu)}$.
Note that $\gh{S_\manM}=n$ and $\p{S_\manM}=n\mod 2$. Master action~\eqref{action-mod} can then be written as
\begin{equation}
\label{action-mod-2}
 S^P=\int d^n x d^n \theta \,\left[ \tilde\Lambda_{A}^{(\mu)} d \tilde\psi^A_{(\mu)}-
\tilde\Lambda_{A}^{(\mu)} \sigma^F \tilde\psi^A_{(\mu)}
+S_{\manM}(\tilde\psi,\tilde\Lambda,x,\theta)\right]\,.
\end{equation}  
The space of field histories can be identified in this representation
with the space of maps from the source supermanifold with coordinates
$x^\mu,\theta^\mu$ into the target-space supermanifold with
coordinates $\psi^A_{(\mu)},\Lambda_A^{(\mu)}$. In particular, the antibracket
\eqref{ab} is induced on the space of maps from the target space bracket~\eqref{brack-M}
(see e.g.~\cite{\GD,Barnich:2009jy} for details on brackets related in this way).

If $\hat L,\gamma$ in~\eqref{action-mod-2} can be chosen $x,\theta$-independent and the second term can be
removed by a field redefinition, then the above master action defines
what is known as the AKSZ sigma model. As we are going to see next
this is exactly what happens if the starting point theory is
diffeomorphism invariant.

\subsection{Diffeomorphism-invariant theories}
\label{sec:diffeom}
We now specialize to the case where the starting point theory is
diffeomorphism invariant and diffeomorphisms are in the generating set
of gauge transformations so that $\gamma$ contains a piece
$\gamma^\prime$ such that $\gamma^\prime \psi^A=(\d_\mu \psi^A)\xi^\mu$, where $\xi^\mu$ are diffeomorphism
ghosts and $\psi^A$ all the fields including $\xi^\mu$.  We assume in addition that this is the only term in $\gamma$
involving undifferentiated $\xi^\mu$.  Under this condition it is
known~\cite{Barnich:1995ap} that by changing coordinates on the space
of local forms as $\xi^\mu -\theta^\mu \to \xi^\mu$, the
$-(\drr_\mu-\dr{x^\mu})\theta^\mu$
term in $\tilde\gamma$ can be
absorbed by $\gamma$ so that
$\tilde\gamma=-\dr{x^\mu}\theta^\mu+\gamma$ after the
redefinition. It then follows that representatives of the
$\tilde\gamma$ cohomology can be assumed $x,\theta$-independent
as we do from now on. Note that in many cases $\hat L$ can be taken
in the form $\xi^1\ldots \xi^n L[\psi]$, where $L$ is a Lagrangian density.

Turning to the parent formulation and following \cite{\BGnlp} we in addition
redefine the $\theta$-descendants of $\xi^\mu$ accordingly, i.e.,
$\xi^\mu_{()\nu} \to \xi^\mu_{()\nu}-\delta^\mu_\nu$ while keeping all the other
fields unchanged.  By this field redefinition, the term $\sigma^F$ in $\gamma^P$
is absorbed into $\bar\gamma$. The following statement follows from
$\tilde\gamma \hat L=0$ and the representation~\eqref{action-mod-2} of the
parent master action
\begin{prop}
  Let the starting point theory be diffeomorphism invariant in the above sense. The function $S_\manM$ defined by~\eqref{Sm} can be then assumed $x,\theta$-independent and hence defines a function on $\manM$ satisfying the following 
master equation
\begin{equation}
\label{ME-M}
 \pb{S_\manM}{S_\manM}_\manM=0\,.
\end{equation} 
Parent master action \eqref{action-mod} can be represented in the explicitly AKSZ form
\begin{equation}
\label{aksz-action}
 S^P=\int d^n x d^n\theta \, \left[\tilde\Lambda^{(\mu)}_A d \tilde \psi^A_{(\mu)}+ S_\manM(\tilde\psi,\tilde\Lambda)\right]\,,
\end{equation}
where the tilde indicates that the variables are now fields depending
on both $x^\mu$ and $\theta^\nu$.
\end{prop}
We stress that in order for~\eqref{aksz-action} to define a theory equivalent
to~\eqref{action-mod}, we need to restrict to field configurations with
$\xi^\mu_{()\nu}(x)$ invertible. Recall also that according to the discussion in
Section~\bref{sec:parent-action} a parent action should be truncated  in order
to be equivalent in a strictly local sense to the starting point action, no
matter which representation is used. It is also worth mentioning that just like
in the non-Lagrangian case considered in~\cite{\BGnlp} once the theory is
rewritten in the form of an AKSZ sigma model one can use generic coordinates
$x^a$ (along with associated $\theta^a$) on the source space that are not at all related to the starting
point coordinates $x^\mu$. Field $\xi^\mu_{()a}(x)$ is then identified as the
respective frame field.

To complete the discussion of the diffeomorphism invariance, we note
that similarly to~\cite{\BGnlp} any theory can be reformulated as an
AKSZ sigma model by adding $y^\mu,\xi^\nu$ as extra variables in the
target space and replacing differential $\bar\gamma$ by its extension $\bar{\tilde\gamma}$ which is $\tilde\gamma$
where the role of $x^\mu,\theta^\nu$ is played by $y^\mu,\xi^\nu$. More precisely, in this case
\begin{equation}
\label{gamma-param}
 \bar{\tilde\gamma} y^\mu=-\xi^\mu\,, \quad  \bar{\tilde\gamma} \xi^\mu=0\,, \quad
\bar{\tilde\gamma} \psi^A_{(\mu_1\ldots \mu_k)}=-\psi^A_{(\mu_1\ldots \mu_{k}\nu)}\xi^\nu+\bar\gamma\psi^A_{(\mu_1\ldots \mu_{k}\nu)}\,.
\end{equation} 
In the Lagrangian setting under consideration now, in addition to extra
coordinates $y^\mu,\xi^\nu$ supermanifold $S_\manM$ also involves their
conjugate antifields/momenta. For instance, in the well-known case of a
1-dimensional system (mechanics) these are the momenta conjugated to time
variable and the reparametrization ghost momenta (see the example in Section~\bref{sec:param}). The master action of the
parametrized parent formulation is then given by \eqref{aksz-action} where in
the expression~\eqref{Sm} for $S_\manM$ differential $\bar\gamma$ is replaced
with the above $\bar{\tilde\gamma}$ and where at the equal footing with $\psi_{(\lambda)}^A$ and their conjugate $\Lambda^{(\lambda)}_A$ the expression involves $y^\mu,\xi^\nu$ and their conjugate antifields/momenta.

We finally comment on the interpretation of the (odd) symplectic
manifold $\manM$ equipped with $\pb{\cdot}{\cdot}$ and $S_\manM$. In
the 1d case the structure of~\eqref{aksz-action} coincides with the
AKSZ-type representation in~\cite{\GD} of the BV master action
associated to a constrained Hamiltonian system with the trivial
Hamiltonian. Moreover, $\manM$ is an extended phase space of the
respective Batalin--Fradkin--Vilkovisky (BFV)
formulation~\cite{Fradkin:1975cq,Batalin:1977pb,Fradkin:1978xi}
with $\pb{\cdot}{\cdot}$ being the extended
Poisson bracket, $S_\manM$ being the BRST charge, and~\eqref{ME-M} the
BFV version of the master equation.  Note that this interpretation is
compatible with the ghost degree and Grassmann parity as
$\gh{S_\manM}=\p{S_\manM}=1$ and the bracket has zero degrees in this
case. Of course, to relate $\manM$ to the usual extended phase space,
one first needs to eliminate many trivial pairs (see, e.g.,~the
example in Section~\bref{sec:particle}). In fact already master
action~\eqref{action-mod-2} can be interpreted in terms of the
Hamiltonian BFV formalism by relating the second term
in~\eqref{action-mod-2} to a Hamiltonian (indeed it can be represented
as a term linear in $\theta^\mu$) in agreement with~\cite{\GD}.  In the
general case it is natural to consider $\manM$ equipped with the
bracket and $S_\manM$ as a multidimensional generalization of the BFV
extended phase space.

\section{Examples}
\subsection{Mechanics}
\label{sec:mechanics}
Consider the  mechanical system described by a Lagrangian $L(q,\d q)$, where $\d$ denotes total time derivative.
If there is no gauge symmetry differential $\gamma$ vanishes and parent
action~\eqref{main} truncated at degree $2$ takes the
familiar form (see e.g.~\cite{Gitman:1990qh})
\begin{equation}
\label{mech-parent}
S^P=S^P_0=\int dt\, \left[p(\d q -q_{(1)})+p^{(1)}(\d q_{(1)}-q_{(2)})+L(q,q_{(1)})\right]\,,
\end{equation} 
where $q_{(l)}=(\dlf{y})^l q$, $p=(\dlf{\theta} q)^*$, and
$p^{(1)}=(\dlf{\theta} q_{(1)})^*$. The total set of variables is
given by $q,q_{(1)},q_{(2)},p,p^{(1)}$, which have zero ghost degree,
and their conjugate in the antibracket variables $q^*,q_{(l)}^*,
l=1,2$ and $\dlf{\theta}q, \dlf{\theta}q_{(1)}$ of ghost degree $-1$.
These last are to be interpreted as antifields. Note that the parent
master action $S^P$ coincides with the classical action $S^P_0$ because there is no gauge
symmetry.

The variables $p,p^{(1)}$ and $q_{(1)},q_{(2)}$ are clearly auxiliary
fields and their elimination brings back the starting point Lagrangian
with $q_{(1)}$ replaced by the ``true'' time derivative
$\d q$. This argument is essentially a specific realization of
the general equivalence proof in Section~\bref{sec:equiv}.

A general feature that can be seen already in this naive example is
that a different reduction is also possible. To see this, we first
eliminate $q_{(2)},p^{(1)}$ as before and suppose for simplicity that
there are no constraints so that equation $p=\dl{q_{(1)}}{L}$ can be
solved for $q_{(1)}$.  The variable $q_{(1)}$ is then an auxiliary
field. Indeed, varying with respect to $q_{(1)}$ gives
$p=\dl{q_{(1)}}L$. Solving this for $q_{(1)}$ gives
\begin{equation}
S_0^\red=\int dt\, (p\d q - (p q_{(1)}(q,p)-L(q,q_{(1)}(q,p)))\,,
\end{equation} 
which is easily recognized as a Hamiltonian action where $p$ plays the role of momenta.
We also note that the respective phase space can be seen as a reduction of the manifold $\manM$
while the canonical Poisson bracket is simply the reduced version of the bracket~\eqref{brack-M}.

This example has a straightforward generalization to the case of field theory without gauge symmetry.
Taking for definiteness the scalar field with Lagrangian $L=\half\d_\mu \phi \d^\mu \phi-V(\phi)$ and
reducing the resulting parent action as in the above example one arrives at
\begin{equation}
\label{p-Ham}
 S_0^{red}=\int d^n x \left[ \pi^\mu\d_\mu \phi -(\half \pi^\mu \pi_\mu+V(\phi))\right]\,.
\end{equation} 
This is a usual first-order action of the scalar field.  We note that
by separating space and time components, this action is seen to become
a Hamiltonian action.

Although the construction is almost trivial in this simple example, it
is much less obvious in the case of gauge theories. From the
perspective of the above example, parent action~\eqref{action-mod} is
a natural generalization of~\eqref{mech-parent} to the case of gauge
field theories. Moreover, this generalization maintains (general)
covariance of the starting point formulation in a manifest way.

We also mention an interpretation of action~\eqref{p-Ham} as a
covariant Hamiltonian action of the De Donder--Weyl (DW) formalism
(see, e.g.,~\cite{Kanatchikov:1997wp,Gotay:1997eg}). For instance the
second term is identified with the DW Hamiltonian while $\pi^\mu$ as
the polymomenta. Moreover, the polysymplectic form
of~\cite{Kanatchikov:1997wp} can be related to the (odd) Poisson bracket~\eqref{brack-M} of
the parent formulation.  A similar interpretation can be given in the
general case and will be discussed elsewhere.

\subsection{Relativistic particle}
\label{sec:particle}

The relativistic particle is defined by the Lagrangian
\begin{equation}
  S[X,\lambda]=\half\int d\tau \big[\lambda^{-1} g_{\mu\nu}(X) \d X^\mu
  \d X^\nu+\lambda m^2\big]=\int d\tau \cL\,.
\end{equation}

The BRST description is achieved by introducing the ghost $\xi$ and the BRST differential
\begin{equation}
\label{s-particle}
  \gamma X^\mu=\xi \d X^\mu\,,\quad \gamma \lambda =\d{(\xi
    \lambda)}\,,\quad
\gamma \xi=\d\xi \xi\,,
  \end{equation}
Note that $\gamma \cL=\d(\xi\cL)$ so that $\hat L=(\xi-\theta)\cL$,
which becomes $\theta$-independent after the redefinition and can be
used in~\eqref{Sm}.

Because of the diffeomorphism invariance, $\gamma\psi^A$ contains $\d\psi^A\xi$
and the parent theory is an AKSZ-type sigma model with the target
space being a supermanifold with the coordinates $X^\mu$, $\xi$,
$\lambda$, all their derivatives $X^\mu_{(l)}$, $\xi_{(l)}$,
$\lambda_{(l)}$ considered as independent coordinates, and canonically
conjugate momenta $p_{\mu}^{(l)},\xi_*^{(l)},\lambda_*^{(l)}$ (these
are momenta not antifields because the bracket~\eqref{brack-M} has
zero ghost degree and Grassmann parity). Here we use the notation such
that $(l)$ refers to the order of the $y$-derivative, e.g.,
$\lambda_{(l)}=(\dlf{y})^l\lambda$. The source space is simply given
by a time line with a coordinate $\tau$ extended by the Grassmann odd
variable $\theta$. The target space function $S_\manM$ is given by
\begin{equation}
 S_\manM=p_\mu \xi {X}_{(1)}^\mu-\xi^*\xi\xi_{(1)}+\lambda^*\xi\lambda_{(1)}+\lambda^*\xi_{(1)}\lambda
+\half\xi(\lambda^{-1}g_{\mu\nu}X^\mu_{(1)} X^\nu_{(1)}+\lambda m^2)+\ldots
\end{equation} 
where dots refer to terms $\Lambda_A^{(l)}\bar\gamma \psi^A_{(l)}$ with $l\geq 1$ and whose explicit form
is in fact not needed here.

It turns out that all the variables except $X,p,\xi,\xi^*$ are trivial
in the sense that all the fields they give rise to (i.e. their
$\theta$-derivatives) are generalized auxiliary fields. By inspecting
the definition of generalized auxiliary fields it follows that it is
enough to show that these variables are generalized auxiliary fields
for $S_\manM$ considered as a master action. In turn, this can be
easily seen using a new coordinate system where $X,\lambda$ are
unchanged while $\xi$ is replaced by $C=\lambda\xi$.  The derivatives
$X^\mu_{(l)},C_{(l)},\lambda_{(l)}$ and conjugate momenta
$p_\mu^{(l)},C_*^{(l)},\lambda_*^{(l)}$ are then defined as before but
starting from the new coordinates and hence are related to the
original ones through a canonical transformation. In terms of the new
coordinate system, $S_\manM$ takes the form
\begin{equation}
 S_\manM=p_\mu \lambda^{-1} C {X}_{(1)}^\mu+\lambda^* C_{(1)}+
\half C (\lambda^{-2}g_{\mu\nu}X^\mu_{(1)} X^\nu_{(1)}+ m^2)+\ldots
\end{equation} 
It is now obvious that $C_{(1)},\lambda-1$ as well as $C_{(n+1)}$,
$\lambda_{(n)}$ for $n\geq 1$, and their conjugate momenta are all
generalized auxiliary fields (we chose $\lambda-1$ because $\lambda$
is assumed invertible).  Moreover, the variables $X^\mu_{(l)}$ and
$p_\mu^{(l)}$ for $l\geq 2$ are also generalized auxiliary fields.

After the elimination we are left with
\begin{equation}
 S^{\rm red}_\manM= C (p_\mu  {X}_{(1)}^\mu+ \half  g_{\mu\nu}X^\mu_{(1)} X^\nu_{(1)}+m^2)\,.
\end{equation} 
In fact ${X}_{(1)}$ and $p^{(1)}$ are also generalized auxiliary
fields because the equation $\ddl{S^{\rm red}_\manM}{X_{1}^\mu}$ can be
algebraically solved for ${X_{1}^\mu}$ ($C$ is to be considered
invertible because it contains an invertible einbein as its $\theta$
descendant). The reduction then gives
 $\Omega=-\half C (g^{\mu\nu}p_\mu p_\nu - m^2)$
which is a BRST charge of the particle model. It is easy to see that the Poisson bracket of the remaining variables is not affected by the reduction\footnote{Strictly speaking the elimination of generalized auxiliary fields
$\chi^a,\chi^*_a$ is the reduction to the second class surface defined by $\chi^*_a=0$ and $\vddl{S}{\chi^a}=0$
so that the reduced bracket is the Dirac bracket (see~\cite{\BGST} for more details).} and is given by
\begin{equation}
 \pb{X^\mu}{p_\nu}_\manM=\delta^\mu_\nu\,, \qquad  \pb{C}{\cP}_\manM=1\,,
\end{equation} 
where we denoted $C_*$ by $\cP$ to agree with the usual conventions of the BFV formalism.

In this way we have reduced the theory to the 1d AKSZ sigma model
with the target space being the BFV phase space of the
relativistic particle equipped with the BRST charge $\Omega$ and the
extended Poisson bracket.  This AKSZ model is known~\cite{\GD} to be
just the BV formulation of the respective first-order Hamiltonian
action.

The example we have just described is the Lagrangian/Hamiltonian version of the
one in~\cite{\BGnlp} (see also~\cite{Brandt:1997iu} for the respective BRST
cohomology treatment).  We stress that although the algebraic procedure that
leads from the Lagrangian to Hamiltonian description of a particle is somewhat
analogous to the usual Legendre transform it is in fact applied to the gauge
theory and is operated in the BRST theory terms. In particular, it allows
identifying constraints and constructing the corresponding BFV--BRST formulation
without actually resorting to the Dirac--Bergmann algorithm and subsequently
constructing the BRST charge.

The last observation in fact remains true in field theory as well.  By
explicitly extracting the ``time'' coordinate and treating the spatial
coordinates implicitly the parent master action can be represented as
a 1d (generalized) AKSZ sigma model of the type proposed in~\cite{\GD}. Its target space comes
equipped with the respective BRST charge and the BRST-invariant
Hamiltonian so that by eliminating the generalized auxiliary fields in
the target space one arrives at the usual BFV description.

\subsection{Parameterized mechanics}
\label{sec:param}
As an example of the parametrized parent Lagrangian let us consider the simplest
and well-known example of a parametrized mechanical system. Starting with the
mechanical system of Section~\bref{sec:mechanics} the parametrization is achieved by
treating the time $t$ as a configuration space coordinate and using new
parameter $\tau$ as a new independent variable. We now construct parametrized
parent formulation as explained at the end of Section~\bref{sec:diffeom}.

Besides the coordinates $q_{(l)}$ and their conjugate
momenta $p^{(l)}$ supermanifold $\manM$ involves in this case coordinate $t$ and reparametrization
ghost $\xi$ along with their conjugate momenta $\pi$, $\cP$. Just like in the previous example 
of relativistic particle bracket~\eqref{brack-M} on $\manM$ is Grassmann even and has vanishing ghost degree.
According to~\eqref{gamma-param} in this case  BRST differential $\bar{\tilde\gamma}$ 
is given by
\begin{equation}
 \bar{\tilde\gamma}t=-\xi\,, \qquad \bar{\tilde\gamma} q_{(l)}=-\xi q_{(l+1)}\,,
\end{equation} 
where we used that $\gamma=\bar\gamma=0$ as the starting point system is not gauge invariant.
Function $S_\manM$ has then the following form:
\begin{equation}
\label{SM-par}
 S_\manM=-\pi\xi-\xi pq_{(1)}
-\xi\sum_{l=1} p^{(l)}q_{(l+1)}
+\xi L_0(q,q_{(1)})\,.
\end{equation}

As in the previous section we take a shortcut and eliminate the auxiliary
variables at the level of $S_\manM$ (of course all the steps can be repeated in
terms of the complete parent master action). Thanks to the third term
in~\eqref{SM-par} variables $p^{(l)},q_{(l+1)}$ for $l>0$ are clearly auxiliary
and can be eliminated without affecting the remaining terms.  Furthermore, if
$L_0(q,q_{(1)})$ is nondegenerate $q_{(1)}$ can be eliminated through its own equation of
motion and one arrives at
\begin{equation}
 S^{\rm red}_\manM=-\xi(\pi+H(q,p))\,,\qquad H=p q_{(1)}(q,p)-L(q,q_{(1)}(q,p))
\end{equation} 
which is easily recognized as the BRST charge implementing the familiar reparametrization
constraint $\pi+H=0$ with the help of reparametrization ghost $\xi$.  Meanwhile the supermanifold
with coordinates $p,q,t,\pi,\xi,\cP$ obtained by reducing $\manM$ is recognized as
the respective BFV phase space. The associated AKSZ action is simply the extended Hamiltonian action implementing
the constraint with the help of the Lagrange multiplier $e=\dlf{\theta}\xi$.
Let us finally emphasize that we have just demonstrated how the parametrized version of the parent formulation
automatically reproduces the Hamiltonian formalism for parametrized systems.

\subsection{Yang--Mills-type theory}
\label{sec:YM}

The set of fields for Yang--Mills-type theory are the components of a Lie
algebra valued 1-form $H_\mu$ and a ghost $C$. The gauge part of the
BRST differential is given by
\begin{equation}
\gamma H_\mu=\d_\mu C+\commut{H_\mu}{C}\,, 
\qquad
\gamma C=\half\commut{C}{C}\,.\\
\end{equation} 
The dynamics is determined by a gauge invariant Lagrangian
$L_0[H]$.

We explicitly identify the field content and the action of the parent
formulation. At ghost number zero we have fields
$(H_\mu)_{(\lambda)[\,]}(x)$ and $C_{(\lambda)|\mu}(x)$. It is useful
to keep the $y$ variables and to work in terms of the following
generating functions:
\begin{equation}
  A_\mu(x|y)=-C_{(\lambda)|\mu}(x)
y^{(\lambda)} \,,\quad B_\mu(x|y)= (H_\mu)_{(\lambda)[\,]}(x)y^{(\lambda)}\,.
\end{equation} 
\comp{
Conventions: $C_\mu=(-1)^{\p{C}}\drf{\theta_\mu} C$, 
$C_{\mu\nu}=C \drf{\theta^\mu}\drf{\theta^\nu}=(-1)^{\p{C}}C_\mu \drf{\theta^\nu}$.
\begin{equation}
d^F C_{\mu\nu}=(-1)^{\p{C}}(-\d_\mu C_\nu+\d_\nu C_\mu),
\end{equation} 
and analogous ones for $\sigma^F\\
$ } The parent action takes the form (for simplicity we keep only
fields of zero ghost number)
\begin{multline}
\label{YM}
S^P_0=\int d^nx \Big[\inner{\pi^{\mu\nu}}{\d_{[\nu} A_{\mu]} -\dl{y^{[\nu}}A_{\mu]}
+\half\commut{A_\nu}{A_\mu}}+\\
\inner{\Pi^{\mu\nu}}{\d_\nu B_\mu-\dl{y^\nu}B_\mu-\dl{y^\mu}A_\nu-\commut{B_\mu}{A_\nu}}+L_0[B]\Big]\,.
\end{multline}
\comp{Conventions: $\pi^{\mu\nu}=(C_{\mu\nu})^*=(C
  \drf{\theta^\mu}\drf{\theta^\nu})^*$ and
  $\Pi^{\mu\nu}=(B_\mu\drf{\theta^\nu})^*$ \\} where we have
introduced the notation
\begin{equation}
\pi^{\mu\nu}(x|p)=\pi^{(\lambda)\mu\nu}(x)p_{(\lambda)}\,,\qquad 
\Pi^{\mu\nu}(x|p)= \Pi^{(\lambda)\mu\nu}(x)p_{(\lambda)}
\end{equation}
for the generating functions containing antifields conjugate to
respectively $C_{(\lambda)|\mu\nu}$ and $(H_\mu)_{(\lambda)|\nu}$.
In addition we introduced inner product $\inner{}{}$ comprising the natural pairing between the Lie
algebra and its dual and the standard inner product (contraction of indices)
between polynomials in $y^\mu$ and $p_\mu$.  The gauge transformation
for all the fields including the Lagrange multipliers $\pi,\Pi$ can be
read off from the complete $S^P$ for which the above $S^P_0$ is the
classical action.  We note that action $S^P_0$ was implicit
in~\cite{Vasiliev:2005zu} (see also~\cite{Vasiliev:2009ck}).  We also mention a somewhat related
formulations in terms of bi-local fields~\cite{Ivanov:1976zq,Witten:1978xx,Ivanov:1979hh}.

Following the same logic as in the above examples, we eliminate
contractible pairs for $-\sigma^F+\bar\gamma$ and their conjugate
antifields. As in~\cite{\BGnlp} it is useful to identify
contractible pairs for $-\sigma^F+\bar\gamma$ as the
$\theta$-descendants of $\tilde\gamma$-trivial pairs in the starting
point jet space. All the jet space coordinates are known to enter
$\tilde \gamma$-trivial pairs except for
$\tilde C=C-\theta^\mu H_\mu$ replacing 
the undifferentiated ghost $C$, curvature 
$F^y_{\mu\nu}=\dlf{y^\mu}H_\nu-\dlf{y^\nu}H_\mu+\commut{H_\mu}{H_\nu}$
and the independent components of its covariant
derivatives. Here we identified jet space coordinates (besides $\theta^\mu,x^\mu$) with the
$y$-derivatives of $C,H_\mu$.  After eliminating the trivial pairs the
reduced differential is determined by the ``Russian formula''
\cite{Stora:1983ct}
\begin{equation}
  \label{YM-RF}
 \tilde\gamma \tilde C=\half[\tilde C, \tilde C]-F^y\,,\qquad F^y=\half F^y_{\mu\nu}\theta^\mu\theta^\nu\,,
\end{equation}
and further relations defining the action of $\tilde\gamma$ on
independent components of (the covariant derivatives of) $F^y_{\mu\nu}$. 

It then follows that all the parent formulation fields are generalized auxiliary except 
the $\theta$-descendants of $\tilde C$ and (the covariant derivatives of) $F^y_{\mu\nu}$ together with
their associated antifields.
Moreover, the action of the reduced $-\sigma^F+\bar\gamma$ can be read off from~\eqref{YM-RF} and its analog for
the curvatures (see~\cite{\BGnlp} for more details). In particular, \eqref{YM-RF} implies
\begin{equation}
 (-\sigma^F+\bar \gamma)^\red \tilde C_{()\mu\nu}=-\commut{\tilde C_{()\mu}}{\tilde C_{()\nu}}+F^y_{\mu\nu}+\ldots
\end{equation} 
where the dots stand for the terms involving fields of nonvanishing
ghost degree.

Assuming that the Lagrangian depends on undifferentiated curvature
only one finds that all the $\theta$-descendants of other curvatures
along with their conjugate antifields are also generalized auxiliary
fields because the corresponding equations of motion merely express
the higher curvatures through the $x$-derivatives of the lower
ones. After eliminating all the above generalized auxiliary fields one
stays with just $\theta$-descendants of $\tilde C$, undifferentiated
curvature $F^y$ and their conjugate antifields. The action for
ghost-number-zero fields $\pi^{\mu\nu}=\half(\tilde C_{()\mu\nu})^*,
A_\mu=-\tilde C_{()\mu},$ and $F^y_{\mu\nu}$ takes then the form
\begin{equation}
\label{YM2}
S_0^\red=\int d^nx \left[\inner{\pi^{\mu\nu}}{\d_\nu A_{\mu}-\d_\mu A_{\nu}+\commut{A_\nu}{A_\mu}-F^y_{\nu\mu}}+L_0(F^y)\right]\,.
\end{equation}
By eliminating $\pi,F$ through their equations of motion one gets the starting point Lagrangian formulation where $F^y$
in $L_0(F^y)$ is replaced with the usual curvature $dA+\half\commut{A}{A}$.

Another reduction of \eqref{YM2} depends on the particular form of
$L_0$.  Taking for definiteness $L_0(F)=
-\frac{1}{4}\eta^{\mu\rho}\eta^{\nu\sigma}
\inner{F_{\mu\nu}}{F_{\rho\sigma}}$ where by slight abuse of
notation $\inner{\,}{\,}$ denotes a nondegenerate invariant form on
the gauge algebra, one observes that varying with respect to $F^y$
allows expressing $F^y$ through $\pi$ as
$F^y_{\mu\nu}=-\eta_{\mu\rho}\eta_{\nu\sigma}\pi^{\rho\sigma}$ where
the identification of the gauge algebra and its dual through the
invariant form is implied. It follows $F^y$ is an auxiliary
field and the reduced action takes the well-known form (see,
e.g.,~\cite{Arnowitt:1962hi})
\begin{equation}
\label{YM3}
S^{\red-1}_0=\int d^n x \big(\inner{\pi^{\mu\nu}}{\dl{x^{\mu}}A_{\nu }-\dl{x^{\nu}}A_{\mu}+\commut{A_\mu}{A_\nu}}+
 \inner{\pi^{\mu\nu}}{\pi_{\mu\nu}}\big)\,.
\end{equation}
We note that the formulation in~\eqref{YM2} has an advantage
over~\eqref{YM3} because it allows for more general Lagrangians, not
necessarily of the form $\inner{F^{\mu\nu}}{F_{\mu\nu}}$. Further generalizations can be achieved
using the parent Lagrangian~\eqref{YM}.

\subsection{Metric Gravity}
\label{sec:grav}

In the BRST description of metric gravity, the fields are the inverse
metric $g^{ab}$ and a ghost field $\xi^a$ that replaces the vector
field parametrizing  an infinitesimal diffeomorphism. The gauge part
of the BRST differential is given by
\begin{equation}
\label{grav-gamma}
  \gamma g^{ab}=L_\xi g^{ab}=\xi^c \d_c g^{ab} -g^{cb}\d_c 
\xi^a -g^{ac}\d_c \xi^b \,,\qquad 
  \gamma \xi^{c}=(\d_a \xi^c) \xi^a\,.  
\end{equation}  
The dynamics is specified by the diffeomorphism-invariant 
Lagrangian $L[g]$ that is assumed to satisfy $\gamma L=\d_a(\xi^a L)$
along with the standard regularity conditions. 

For metric gravity, $\gamma X$ contains $(\d_a X) \xi^a$ for any field
$X$ so that the general discussion of diffeomorphism-invariant
theories applies.  In particular, $\hat L$ representing the Lagrangian 
can be chosen as $\hat L=\xi^0\ldots\xi^{n-1} L_0[g]$ and parent
formulation can be written as the AKSZ sigma model. Its target
space has coordinates $\xi^a_{(b)}$, $g^{ab}_{(c)}$ along with their
canonically conjugate antifields/momenta $\pi_a^{(b)}$ and
$u_{ab}^{(c)}$.

It is useful to work in terms of generating functions. For this, we
introduce formal variables $p_b$ in addition to $y^a$ and consider the
algebra of polynomials in $y,p$ equipped with the standard Poisson
bracket $\pb{y^a}{p_b}=\delta_a^b$.
The target space coordinates
$g^{ab}_{c_1\ldots c_l}$ and $ \xi^a_{c_1\ldots c_l}$ can then
be encoded in
\begin{equation}
G=\half g^{ab}_{(c)}y^{(c)}p_ap_b\,, \qquad \Xi= \xi^{a}_{(c)}y^{(c)}p_a\,,
\end{equation} 
and the action of $\gamma$ on these coordinates can be
compactly written as 
\begin{equation}
 \gamma \Xi=\half \{\Xi,\Xi\}\,,  \qquad  
\gamma G=\pb{G}{\Xi}\,.\label{eq:liegrav}
\end{equation}

The same variables can be used to encode antifields/momenta into the generating functions:
\begin{equation}
 \Pi=\pi_a^{(b)}p_{(b)}y^a\,, \qquad U=\half u_{ab}^{(c)}p_{(c)}y^ay^b\,.
\end{equation} 
In addition, we introduce the natural symmetric inner product
$\inner{}{}$ on the space of polynomials in $y,p$ such that e.g.~
$\inner{y^a}{p_b}=\delta^a_b$. In components it simply amounts to
natural contraction between indices of the coefficients. The parent
master action then becomes
\begin{equation}
\begin{gathered}
 S^P=\int d^nx d^n \theta\left[ \inner{\tilde U}{d^F \tilde G}+\inner{\tilde \Pi}{d^F\tilde \Xi}
+ S_\manM(\tilde G,\tilde \Xi,\tilde U,\tilde \Pi)\right]\,,\\[5pt]
 S_\manM=\inner{\tilde U}{\pb{\tilde G}{\tilde \Xi}}+\half\inner{\tilde \Pi}{\pb{\tilde \Xi}{\tilde \Xi}}+\tilde\xi^0\ldots\tilde\xi^{n-1} L_0[\tilde G]\,.
 \end{gathered}
\end{equation}
where $\tilde\xi$ enters $\tilde\Xi$ as a $y$-independent term and where as before the tilde indicates that the fields are functions of $x,\theta$.

We now concentrate on the classical action~$S^P_0$. 
Fields $F,A$ of vanishing ghost degree enter the expansions of $G,\Xi$ in
$\theta$ as
\begin{equation}
\tilde G(x,\theta|y,p)=F(x,y,p)+\ldots \,, \quad \tilde \Xi(x,\theta|y,p)=\Xi(x|y,p)+A_\mu(x|y,p)\theta^\mu+\ldots\,.
\end{equation} 
As regards the antifields/momenta, the
$n-1$-form $P$ and $n-2$ form $\pi$ components of respectively  $U$ and $\Pi$ are of vanishing ghost
degree and play the role of Lagrange multipliers. The classical action can be then written as
\begin{equation}
\label{grav-class-action}
 S^P_0=\int d^nx d^n \theta \big[ \inner{P}{d F +\pb{F}{A}}+\inner{\pi}{dA +\half\pb{A}{A}}
+e^0...\,e^{n-1} L_0[F]\big]\,,
\end{equation} 
where $e^a=e^a_\mu(x)\theta^\mu$ enters $A(x,\theta|y,p)$ as $A=\theta^\mu e_\mu^a(x) p_a+\ldots$ and is to be identified
as the frame field. Action~\eqref{grav-class-action} was implicitly in~\cite{Vasiliev:2005zu} (see also~\cite{Vasiliev:2009ck}).
We also mention somewhat related descriptions from~\cite{Borisov:1974bn,Pashnev:1997xk}.

We now perform the reduction of the parent formulation for gravity
leading to its frame like form. We are going to implement the
Lagrangian version of the analogous reduction considered
in~\cite{Barnich:2010sw} (see
also~\cite{Vasiliev:2005zu,\BGST}). Details on identification of
trivial pairs for the BRST differential can be found
in~\cite{Brandt:1996mh,Barnich:1995db,Barnich:1995ap}. In particular,
all the variables in $\Xi$ and $G$ except $\xi^a_{()},\xi^a_{b}$,
metric $g^{ab}$, and (the independent components of the covariant
derivatives of) the curvature are contractible pairs for
$\bar\gamma$. All their $\theta$-descendants as well as all the
associated antifields are then the generalized auxiliary fields for
the parent formulation.  Moreover, under the usual assumption that
metric (entering $G$ as a $g^{ab}p_ap_b$) is close to a flat metric
$\eta^{ab}$, the components of the difference $g^{ab}-\eta^{ab}$
together with the symmetric part of $\xi^a_c\eta^{cb}$ and their
associated antifields give rise to generalized auxiliary fields and
hence can also be eliminated.

The action of the reduced $\bar\gamma$ on the remaining coordinates 
$\xi^a$, $\xi^a_b$, $R^a_{b\, cd}$ and $R^b_{c_1 \ldots c_k a_1 a_2 a_3}$, where the latter denote the covariant derivatives 
of the curvature $R^a_{b\, cd}$ is given by (see e.g.~\cite{\BGnlp,Brandt:1996mh,Barnich:1995ap} for more details)
\begin{equation}
  \bar\gamma \xi^a=\xi^a_c\xi^c\,,\qquad \bar\gamma \xi^{a}_b=\xi^a_c\xi^c_b-\half \xi^c\xi^d R^a_{b\, cd}\,,\qquad \\
\end{equation}
and
\begin{multline}
\label{gamma-R}
 \bar\gamma 
  R^b_{c_1\ldots c_ka_1a_2a_3}=\xi^{c_0}
  R^b_{c_0c_1 \ldots c_k a_1a_2a_3} -\xi^b_d 
  R^d_{c_1\dots c_k a_1a_2a_3}+\\
+\xi^d_{c_1} R^b_{d c_k a_1a_2a_3}+\dots + \xi^d_{a_3}
  R^b_{c_1\ldots c_k a_1a_2d}\,.
\end{multline} 
If $L_0$ depend on undifferentiated curvature only 
all the fields associated to the covariant derivatives of the curvature are generalized auxiliary. Indeed,
it follows from~\eqref{gamma-R} that the respective equations of motion express $R^b_{c_1 \ldots c_k a_1 a_2 a_3}$ through $R^b_{c_1 \ldots c_{k-1} a_1 a_2 a_3}$ so that $\theta$-derivatives of $R^b_{c_1 \ldots c_{k} a_1 a_2 a_3}$ with $k>0$
and all the associated antifields can be eliminated. In this way one ends up with only $\theta$ derivatives
of $\xi^a,\xi^a_b,R^a_{b,cd}$ and the associated antifields/momenta.

We then introduce the component fields entering
$\tilde\xi^a,\tilde\xi^a_b,\tilde R^a_{b,cd}$:
\begin{equation}
\begin{gathered}
 \tilde\xi^a(x,\theta)=\xi^a-\theta^\mu e_\mu^a+\half\theta^\nu\theta^\mu\xi^a_{\mu\nu}+\ldots\,, \\
 \tilde\xi^a_b(x,\theta)=\xi^a_b-\theta^\mu \omega^a_{\mu b}+\half\theta^\nu\theta^\mu\xi^a_{b \mu\nu}+\ldots\,, \quad
 \tilde R^a_{b,cd}(x,\theta)=R^a_{b,cd}+\ldots
\end{gathered}
\end{equation} 
where dots stand for terms of higher order in $\theta$. In particular, fields
$e^a_\mu,\omega^a_{\mu b},R^a_{b,cd}$ carry vanishing ghost degree. Besides them
antifields $\pi^{\mu\nu}_a=(\xi^a_{\mu\nu})^*$ and $\pi^{b\mu\nu}_a=(\xi^a_{b \mu\nu})^*$ 
also carry vanishing ghost degree and play the role of Lagrange multipliers. After the reduction action~\eqref{grav-class-action} takes the following form
\begin{multline}
\label{grav-interm}
 S^{\red}_0[\pi^a,\pi^a_b,e^a,\omega^{ab}]=\int d^n x d^n \theta
\Big[\pi_{a}(de^a+\omega^a_b e^b)
+\\
+\pi^{b}_a
(d\omega_{b}^a +\omega_{c}^a\omega^c_{b}-\half e^c e^d R^a_{bcd})+
e^0\ldots e^{n-1} L_0(R)\Big]\,,
\end{multline} 
where antifields $\pi_a$ and $\pi_a^b$ are represented in a dual way as $n-2$-forms.
Fields $\pi^{b\mu\nu}_a$ and $R^a_{bcd}$ are clearly auxiliary ones. By eliminating them the second term is gone
and we get 
\begin{equation}
\label{grav-interm-2}
S^{\red-1}_0[\pi^a,e^a,\omega^{ab}]=\int d^n x 
\pi^{\mu\nu}_{a}
(\d_{[\nu} e^a_{\mu]}+\omega^{ac}_{[\nu} e^c_{\mu]})
+ \int d^nx d^n\theta\,\,e^0\ldots e^{n-1} L_0[e,\omega]\,.
\end{equation}

Just like in other examples, it is now easy to explicitly get back the
starting point Lagrangian. Indeed, the fields $\pi$ and $\omega$ are
auxiliary because varying with respect to $\pi^{\mu\nu}_a$ gives the
condition $de^a+\omega^a_b e^b=0$ that is uniquely solved for
$\omega^a_b$ in terms of $e^a$. At the same time variation with
respect to $\omega^a_{\mu b}$ gives equation $\pi^{\mu\nu}_a
e^b_\nu+(\text{$\pi$ -independent terms})=0$ which can be uniquely
solved for $\pi^{\mu\nu}_a$. Substituting the solutions back to~\eqref{grav-interm-2}
one finds that only the term with $L_0$ expressed through $e^a$ stays.

If the starting point $L_0$ is precisely
the Einstein-Hilbert Lagrangian another reduction is also possible
that leads to the usual first order action
\begin{equation}
\label{palatini}
S_1[e^a,\omega^{ab}]=\int d^nx d^n\theta \,\, \epsilon_{a_1\ldots a_{n-2}a_{n-1}a_{n}}e^{a_1}\ldots e^{a_{n-2}}(d\omega^{a_{n-1}a_{n}}+
\omega^{a_{n-1}}_c\omega^{c a_{n}})\,,
\end{equation} 
depending on $e^a,\omega^a_b$ as independent fields. The difference with~\eqref{grav-interm-2}
is only in the first term in~~\eqref{grav-interm-2} and its extra dependence on $\pi^{\mu\nu}_a$.
That~\eqref{grav-interm-2} is equivalent to~\eqref{palatini} via eliminating auxiliary fields is obvious
if one eliminates $\pi^{\mu\nu}_a$ and $\omega^{ab}_\mu$ in~\eqref{grav-interm-2} as explained above
and eliminates $\omega^{ab}_\mu$ through its own equations of motion in~\eqref{palatini}.

In fact~\eqref{palatini} can be obtained from~\eqref{grav-interm-2} via a straightforward reduction. Indeed,
let us change the field variables such that $\omega^{ab}_\mu=\alpha^{ab}_\mu(e)+\bar\omega^{ab}_\mu$ where $\alpha^{ab}_\mu[e]$ is a unique solution to $de^a+\alpha^a_c e^c=0$ so that field $\bar\omega^{ab}_\mu$ is related to torsion in an invertible way. In terms of $\bar\omega^{ab}_\mu$ action~\eqref{palatini} decomposes as $S_1[e,\alpha(e)]+S_2[e,\bar\omega]$ where $S_2$ is bilinear in undifferentiated $\bar \omega^{ab}_\mu$. Using this representation for the second term in~\eqref{grav-interm-2} one observes that $\bar\omega^{ab}_\mu$ is an auxiliary field and can be expressed through $\pi^{\mu\nu}_a$ and $e^a_\mu$. Using then an invertible
field redefinition such that a new $\omega^{ab}_\mu[e,\pi]$ replaces $\pi^{\mu\nu}_a$ the reduced action~\eqref{grav-interm-2} can be brought to the form~\eqref{palatini}.

\section{Conclusions}
In this paper, we have specialized the parent formulation of~\cite{\BGnlp} to
the Lagrangian level.  More precisely, for a given Lagrangian gauge theory, we
have constructed the first-order parent BV formulation by explicitly specifying
the field--antifield space, the antibracket, and the BV master action.  As a
technical assumption, we restricted ourselves to the case of theories with a
closed gauge algebra.  But the parent formulation can also be defined in
general. Indeed, $S^P$ can be defined in exactly the same way, and the only
difference is that in the general case, it satisfies the master equation only
modulo the parent equations of motion. These last are determined by the
classical action $S^P_0$, which is also well defined in general and can be
obtained from $S^P$ by putting all the fields of nonzero ghost degrees to zero.
The complete master action can then be obtained via the usual BV procedure
starting from $S^P_0$ and its gauge symmetries.

Although the construction of the parent formulation applies to an already
specified gauge theory, our hope is to use this formulation to construct new
models in the parent form (or related forms) from the very beginning.  This
strategy has proved fruitful~\cite{\AGT,Bekaert:2009fg,Alkalaev:2009vm} in the
context of higher-spin gauge theories, where a version of the parent formulation
at the level of the equations of motion~\cite{\BGST,\BGnlp,Barnich:2005ru}  was
successfully used.  

Among possible applications of the present results, Vasiliev's interacting
higher-spin theory~\cite{Vasiliev:1988sa,Vasiliev:1992av,Vasiliev:2003ev}, where
the Lagrangian formulation is currently unknown, seems to be the most
attracting. We hope that the present approach gives the correct framework for
addressing this issue. This is supported by a concise parent-like formulation of
the nonlinear higher-spin theory at the off-shell level~\cite{\Goff} (see
also~\cite{Vasiliev:2005zu}). As far as higher spin fields are concerned let us
note that the present approach should give a systematic way to derive frame-like
actions (such as those
of~\cite{Vasiliev:1980as,Lopatin:1988hz,Skvortsov:2008sh}) starting from the
metric-like ones or provide a framework for addressing this problem for systems
where Lagrangian formulation in not available such as, e.g., mixed symmetry AdS
fields where actions are known only for particular
cases~\cite{Zinoviev:2003dd,Alkalaev:2003hc,Alkalaev:2003qv,Zinoviev:2009gh}.
Another interesting perspective is to relate the parent action to that of the
recently proposed double field theory \cite{Hull:2009mi,Hohm:2010jy}.
\section*{Acknowledgments}
\label{sec:acknowledgements}

\addcontentsline{toc}{section}{Acknowledgments}

The author is grateful to G.~Barnich for the collaboration at an early
stage of this project.  He wishes to thank K.~Alkalaev, I.~Batalin,
A.~Semikhatov, E.~Skvortsov, D.~Roytenberg, R.~Metsaev, I.~Tyutin, and
M.~Vasiliev for discussions. This work has been partially completed
during a stay at the Erwin Schr\"odinger International Institute for
Mathematical Physics and supported by the RFBR grant 10-01-00408-a and
the RFBR-CNRS grant 09-01-93105

\addtolength{\baselineskip}{-3pt}
\addtolength{\parskip}{-3pt}

\providecommand{\href}[2]{#2}\begingroup\raggedright\endgroup

\end{document}